\documentclass[aps,onecolumn,nofootinbib,tightenlines,superscriptaddress]{revtex4-2}

\usepackage{amssymb,amstext,amsmath,amsthm,slashed}
\usepackage[dvips]{graphicx}
\usepackage{latexsym}
\usepackage{psfrag}
\usepackage{amsfonts}
\usepackage{color}
\usepackage{verbatim}
\usepackage{bbm}
\usepackage{dcolumn}
\usepackage{tabularx}
\usepackage{leftidx}
\usepackage{tensor}
\usepackage{boxedminipage}
\usepackage{fancyvrb}
\usepackage{float}
\usepackage{wrapfig}
\usepackage{booktabs}
\usepackage{multirow}
\usepackage{subfigure}
\usepackage{dsfont} 
\usepackage{mathrsfs}
\usepackage[hidelinks]{hyperref}
\usepackage{scalerel,stackengine}
\usepackage{empheq}

\begin{document}

\title{An Exact, Time-dependent Analytical Solution for the Magnetic Field \\ in the Inner Heliosheath \vspace{0.1cm}}


\author{Christian R\"oken \vspace{0.25cm}} 

\email{croeken@uni-bonn.de}

\affiliation{Lichtenberg Group for History and Philosophy of Physics - Faculty of Philosophy, University of Bonn, 53113 Bonn, Germany \vspace{0.25cm}}

\affiliation{Department of Geometry and Topology - Faculty of Science, University of Granada, 18071 Granada, Spain \vspace{0.25cm}}


\author{Jens Kleimann} 

\email{jk@tp4.rub.de}

\affiliation{Institute for Theoretical Physics IV - Faculty of Physics and Astronomy, Ruhr University Bochum, 44780 Bochum, Germany \vspace{0.25cm}}


\author{Horst Fichtner} 

\email{hf@tp4.rub.de}

\affiliation{Institute for Theoretical Physics IV - Faculty of Physics and Astronomy, Ruhr University Bochum, 44780 Bochum, Germany \vspace{0.25cm}}

\affiliation{Research Department Plasmas with Complex Interactions, Ruhr University Bochum, 44780 Bochum, Germany\vspace{0.4cm}}


\date{October 2021 / December 2022}

\begin{abstract}
\vspace{0.4cm} \noindent \textbf{\footnotesize ABSTRACT.} \, We derive an exact, time-dependent analytical magnetic field solution for the inner heliosheath, which satisfies both the induction equation of ideal magnetohydrodynamics in the limit of infinite electric conductivity and the magnetic divergence constraint. To this end, we assume that the magnetic field is frozen into a plasma flow resembling the characteristic interaction of the solar wind with the local interstellar medium. Furthermore, we make use of the ideal Ohm's law for the magnetic vector potential and the electric scalar potential. By employing a suitable gauge condition that relates the potentials and working with a characteristic coordinate representation, we thus obtain an inhomogeneous first-order system of ordinary differential equations for the magnetic vector potential. Then, using the general solution of this system, we compute the magnetic field via the magnetic curl relation. Finally, we analyze the well-posedness of the corresponding Dirichlet-type initial-boundary value problem, specify compatibility conditions for the initial-boundary values, and outline the implementation of initial-boundary conditions. 
\end{abstract}

\setcounter{tocdepth}{2}

\vspace{0.1cm}

\maketitle

\tableofcontents

\section{Introduction} \label{SecI}

\noindent The outer heliosphere has been in the focus of various research activities in recent years. On the one hand, it has long been recognized as a region of significance for the modulation of cosmic rays (see, e.g., \cite{Potgieter-2013}), the acceleration of energetic particles (for instance, \cite{Richardson-2013}), the filtration of neutral atoms out of the interstellar gas penetrating into the heliosphere \cite{Moebius-etal-2009b, Bzowski-etal-2019}, or the reconnection of the heliospheric with the local interstellar magnetic field \cite{Opher-etal-2017, Pogorelov-etal-2017a}. On the other hand, with the entry of the two Voyager spacecraft into the inner heliosheath, i.e., the region enclosed by the heliopause and the termination shock, the in-situ exploration of the outer heliosphere has begun. Recently, both spacecraft have traversed the inner heliosheath in full: After Voyager~1 crossed the heliopause in April 2013, Voyager~2 did so in November 2018. Their in-situ measurements are supplemented with remote observations made with the Interstellar Boundary Explorer. Additionally, a new mission termed \textit{Interstellar Mapping and Acceleration Probe} will be launched in 2024 \cite{McComas-etal-2018}. A comprehensive summary of both the present knowledge about the outer heliosphere and its astrophysical significance can be found in the review paper \cite{Pogorelov-etal-2017b}.

For several of the above topics, knowledge about the large-scale spatial structure of the magnetic field in the heliosphere and its outer vicinity, as well as its time dependence in the inner heliosheath is of particular interest. While quantitatively `precise' treatments will eventually result from large-scale numerical models (like those in \cite{Opher-etal-2017,Washimi-etal-2017, Izmodenov-2018, Gamayunov-etal-2019, Heerikhuisen-etal-2019}), for some questions the use of simplifying analytical models is still desirable (see, e.g., \cite{Schwadron-etal-2014, Isenberg-etal-2015, Sylla-Fichtner-2015}). For the (draped) local interstellar magnetic field, we have thus presented analytical solutions in a series of three papers. The first fully analytical work of this type was the model in \cite{Roeken-etal-2015}, where we derived an exact solution for the magnetic field frozen into the incompressible Rankine half-body flow that was originally formulated by Parker in \cite{Parker-1961} as a heliospheric flow pattern. A similar solution to this problem, which however suffers from the presence of coordinate singularities and relies on certain approximation schemes, was given by Isenberg et al.\ in the almost contemporaneous paper \cite{Isenberg-etal-2015}. Moreover, as a common shortcoming of all analytical [magneto]hydrodynamical models of the large-scale heliosphere and its interaction with the local interstellar medium is a heliotail exhibiting a circular cross section although both numerical simulations (for instance, \cite{Heerikhuisen-etal-2014}) as well as data from the Interstellar Boundary Explorer \cite{McComas-etal-2013} confirm a strong flattening of the heliotail by the interstellar magnetic field, we have removed this inconsistency in \cite{Kleimann-etal-2016} using the method of \textit{distortion flows}. Finally, in \cite{Kleimann-etal-2017}, we generalized the model introduced in \cite{Roeken-etal-2015} to compressible flows, sacrificing neither generality nor tractability.

With the present paper, we transfer the first part of our analytical treatment of the local interstellar magnetic field to the heliospheric magnetic field in the inner heliosheath. Early analytical models in this direction date back to the work \cite{Yu-1974} of Yu in 1974, who applied the Cauchy integral formalism \cite{Cauchy-1816} to describe the advection of a magnetic field in the flow from the shock surface into the wake of the solar wind, and deduced a field topology of two spiral tubes confined within a single heliotail, albeit only in the limit of very large tailward distances. Similar potential flow fields of constant density have also been employed in \cite{Nerney_Suess-1995, Senanayake-Florinski-2013, Nicolaou-Livadiotis-2017} to determine termination shock geometries deviating from that of a sphere and to study the resulting flow properties. Furthermore, in \cite{Drake-etal-2015}, magnetohydrodynamical estimates obtained by neglecting the flow of the local interstellar medium indicated the Parker spiral field's ability to drive so-called polar jets. Still missing, however, is not only an analytical expression for the magnetic field in the inner heliosheath that is consistent with the (steady-state) plasma flow used in our previous papers and those by some of the aforementioned authors but also one that takes into account the inherent time dependence of the inner heliosheath as a consequence of the solar cycle. Therefore, we here derive a general analytical magnetic field solution that is both consistent with the potential flow and time-dependent.

The paper is organized as follows. After some preliminary considerations in Section \ref{SecII}, we present our magnetic field solution for the inner heliosheath in Section \ref{SecIII}. Initial-boundary conditions are the subject of Section \ref{SecIV}. Finally, in Section \ref{SecV}, we conclude with a summary of our findings and a brief outlook toward a future research project.

\section{Preliminaries} \label{SecII}

\noindent We recall the relevant basics of ideal magnetohydrodynamics, beginning with a specification of the geometric setting. We work on an open, unbounded, and connected proper subset $\Omega \subset \mathfrak{M} \cong \mathbb{R}^3$ of a Riemannian $3$-manifold $(\mathfrak{M}, \boldsymbol{\mathcal{E}})$ with Euclidean metric $\boldsymbol{\mathcal{E}} = \delta_{i j} \, \textnormal{d}x^i \otimes \textnormal{d}x^j$, $i, j \in \{1, 2, 3\}$, and points $\boldsymbol{x} = (X, Y, Z) \in \mathfrak{M}$ in a Cartesian coordinate representation. Moreover, we assume the boundary $\overline{\Omega} \setminus \Omega =: \partial \Omega = \partial \Omega_{\textnormal{in}} \cup \partial \Omega_{\textnormal{out}}$ to consist of an inner component $\partial \Omega_{\textnormal{in}} \cong S^2$ that is smooth and compact without boundary and an outer component $\partial \Omega_{\textnormal{out}} \cong \mathbb{R}^2$ being smooth and closed without boundary. In order to characterize physical as well as gauge quantities at each instant of time, we introduce the parameter $t \in I \subseteq \mathbb{R}_{\geq 0}$ and define the metric on $I \times \overline{\Omega}$ by the Minkowski metric $\boldsymbol{\eta}$ with signature $(3, 1, 0)$. As in \cite{Roeken-etal-2015}, we use the incompressible and irrotational Rankine-type flow field $\boldsymbol{u} \hspace{-0.07cm} : I \times \overline{\Omega} \rightarrow \mathbb{R}^3$ given by
\begin{equation} \label{flowfield}
\boldsymbol{u} = u_0 \biggl[\frac{q \rho}{(\rho^2 + z^2)^{3/2}} \, \boldsymbol{e}_{\rho} + \biggl(\frac{q z}{(\rho^2 + z^2)^{3/2}} - 1\biggr) \boldsymbol{e}_z\biggr] \, ,
\end{equation}
where the transformation 
\begin{equation*}
\mathfrak{T}^{(1)} \hspace{-0.05cm}:
\begin{cases}
\, \mathbb{R} \times \mathbb{R} \times \mathbb{R} \rightarrow \mathbb{R}_{> 0} \times \mathbb{R} \times [0, 2 \pi) \vspace{0.2cm} \\ 
\hspace{0.30cm} (X, Y, Z) \mapsto (\rho, z, \varphi)
\end{cases} 
\end{equation*}
with
\begin{equation} \label{T0C} 
\rho = \sqrt{X^2 + Y^2} \, , \quad z = Z \, , \quad \textnormal{and} \quad \varphi = \textnormal{sgn}(Y) \biggl[\arccos{\biggl(\frac{X}{\sqrt{X^2 + Y^2}}\biggr)} - \pi\biggr] + \pi
\end{equation}
yields the mapping from the Cartesian to a cylindrical coordinate representation, $u_0$ is the speed of the homogeneous interstellar flow at infinity, and $q \in \mathbb{R}_{> 0}$ is the relative strength of the stationary point-like solar wind source located at the origin. Besides, we employ the convention $\textnormal{sgn}(0) := 1$. To obtain the magnetic field $\boldsymbol{B} \hspace{-0.07cm} : I \times \overline{\Omega} \rightarrow \mathbb{R}^3$, we have to solve the Dirichlet-type initial-boundary value problem 
\begin{empheq}[left={\empheqlbrace}]{alignat=2}
& \boldsymbol{\nabla} \times (\boldsymbol{u} \times \boldsymbol{B}) = \partial_t \boldsymbol{B} \quad \textnormal{in} \,\, I \times \Omega \label{induction} \\[0.2cm] 
& \hspace{1.58cm} \boldsymbol{\nabla} \cdot \boldsymbol{B} = 0 \quad \textnormal{in} \,\, I \times \Omega \label{mdc} \\[0.2cm] 
& \hspace{0.89cm} \boldsymbol{B}_{|\{t_{\textnormal{B}}\} \times \partial \Omega} = \boldsymbol{g} \quad \textnormal{on} \,\, \{t_{\textnormal{B}}\} \times \partial \Omega \, , \label{BV}
\end{empheq}
which consists, top down, of the induction equation in the limit of infinite electric conductivity, the magnetic divergence constraint, and an initial-boundary condition with Dirichlet-type initial-boundary values $\boldsymbol{g} \in C^0\bigl(\{t_{\textnormal{B}}\} \times \partial \Omega, \mathbb{R}^3\bigr)$, where $t_{\textnormal{B}}$ is the initial time value on the boundary $\partial \Omega$. We point out that in Section \ref{SecIIIA}, we identify the outer boundary $\partial \Omega_{\textnormal{out}}$ with the heliopause, i.e., with the separatrix of the streamlines of the flow field (\ref{flowfield}). Accordingly, we impose boundary values only at the inner boundary $\partial \Omega_{\textnormal{in}} \equiv \partial \Omega$. We further remark that by applying the vector calculus identity for the curl of a cross product, the incompressibility condition $\boldsymbol{\nabla} \cdot \boldsymbol{u} = 0$ for the flow field, as well as the magnetic divergence constraint, the above limit of the induction equation simplifies to  
\begin{equation} \label{redineq}
(\boldsymbol{B} \cdot \boldsymbol{\nabla}) \boldsymbol{u} - (\boldsymbol{u} \cdot \boldsymbol{\nabla}) \boldsymbol{B} = \partial_t \boldsymbol{B} \, .
\end{equation}
Instead of solving this initial-boundary value problem, we may alternatively consider the initial-boundary value problem
\begin{empheq}[left={\empheqlbrace}]{alignat=2}
& \boldsymbol{u} \times (\boldsymbol{\nabla} \times \boldsymbol{A}) = \partial_t \boldsymbol{A} + \boldsymbol{\nabla} \psi \quad \textnormal{in} \,\, I \times \Omega \label{tdiol} \\[0.2cm] 
& \hspace{2.25cm} \boldsymbol{B} = \boldsymbol{\nabla} \times \boldsymbol{A} \quad \textnormal{in} \,\, I \times \Omega \label{BArel} \\[0.2cm] 
& \hspace{1.88cm} \boldsymbol{B}_{|\{t_{\textnormal{B}}\} \times \partial \Omega} = \boldsymbol{g} \quad \textnormal{on} \,\, \{t_{\textnormal{B}}\} \times \partial \Omega \, , \nonumber
\end{empheq}
where equation (\ref{tdiol}) is the ideal Ohm's law and equation (\ref{BArel}) the magnetic curl relation with $\boldsymbol{A} \hspace{-0.07cm} : I \times \overline{\Omega} \rightarrow \mathbb{R}^3$ being the magnetic vector potential and $\psi \hspace{-0.07cm} : I \times \overline{\Omega} \rightarrow \mathbb{R}$ the electric scalar potential. The equivalency between these two initial-boundary value problems can be easily seen as follows. We substitute the magnetic curl relation (\ref{BArel}) into the limit (\ref{induction}) of the induction equation, which yields
\begin{equation*}
\boldsymbol{\nabla} \times [\boldsymbol{u} \times (\boldsymbol{\nabla} \times \boldsymbol{A})] = \boldsymbol{\nabla} \times \partial_t \boldsymbol{A} \, .
\end{equation*}
Then, employing the Poincar\'e lemma, we immediately find the ideal Ohm's law (\ref{tdiol}). Moreover, the curl relation satisfies the magnetic divergence constraint (\ref{mdc}) trivially. This proves the claim. Also, by using the vector calculus identity for the gradient of a scalar product and the irrotationality condition $\boldsymbol{\nabla} \times \boldsymbol{u} = \boldsymbol{0}$ for the flow field, the ideal Ohm's law reduces to the form
\begin{equation} \label{OhmSimp} 
\boldsymbol{\nabla} (\boldsymbol{u} \cdot \boldsymbol{A}) - (\boldsymbol{u} \cdot \boldsymbol{\nabla}) \boldsymbol{A} - (\boldsymbol{A} \cdot \boldsymbol{\nabla}) \boldsymbol{u} = \partial_t \boldsymbol{A} + \boldsymbol{\nabla} \psi \, .
\end{equation}
In Section \ref{SecIIIB}, we subject this equation to a specific axial gauge relating the potentials, and determine a general solution $\boldsymbol{A} \in C^2(I \times \Omega, \mathbb{R}^3)$. Afterwards, we compute the magnetic field via the curl relation (\ref{BArel}). Particular solutions $\boldsymbol{B} \in C^1(I \times \Omega, \mathbb{R}^3) \cap C^0(I \times \overline{\Omega}, \mathbb{R}^3)$ for the magnetic field will be discussed in a separate paper.

\section{Derivation of an Exact, Time-dependent Analytical Solution for the Magnetic Field in the Inner Heliosheath} \label{SecIII}

\subsection{Domain of the Inner Heliosheath and Characteristic Coordinates} \label{SecIIIA}

\noindent To define the domain of the inner heliosheath, we make use of the implicit surface equation for its outer boundary, the heliopause, which in a cylindrical coordinate representation reads
\begin{equation*}
2 q - \rho^2 - z(\rho) \, \sqrt{4 q - \rho^2} = 0 \quad \textnormal{with} \quad \rho < 2 \sqrt{q} \, .
\end{equation*}
This equation gives rise to upper bounds in the cylindrical images
\begin{equation*}
\textnormal{Ran}(\rho) = (0, 2 \sqrt{q}) \quad \textnormal{and} \quad \textnormal{Ran}(z) = \Biggl(- \infty, \frac{2 q - \rho^2}{\sqrt{4 q - \rho^2}}\Biggr] \, .
\end{equation*}
Furthermore, we identify the inner boundary $\partial \Omega$ with the termination shock. Thus, carrying out the transformation from cylindrical to spherical coordinates 
\begin{equation*} 
\mathfrak{T}^{(2)} \hspace{-0.05cm}:
\begin{cases}
\, \displaystyle (0, 2 \sqrt{q}) \times \Biggl(- \infty, \frac{2 q - \rho^2}{\sqrt{4 q - \rho^2}}\Biggr] \times [0, 2 \pi) \rightarrow \Biggl(0, \sqrt{\frac{2 q}{\smash{1 + \sin{(\vartheta)}}}} \,\, \Biggr] \times \biggl(- \frac{\pi}{2}, \frac{\pi}{2}\biggr) \times [0, 2 \pi) \\ \\
\hspace{4.75cm} (\rho, z, \varphi) \mapsto (r, \vartheta, \phi)
\end{cases} 
\end{equation*}
with
\begin{equation} \label{SphericalCR} 
r = \sqrt{\rho^2 + z^2} \, , \quad \vartheta = \arctan{\biggl(\frac{z}{\rho}\biggr)} \, , \quad \textnormal{and} \quad \phi = \varphi \, ,
\end{equation}
and employing the implicit surface equation for $\partial \Omega$
\begin{equation} \label{boundarypar}
r - R(\vartheta, \phi) = 0 \, , 
\end{equation}
where the function
\begin{equation*} 
R(\vartheta, \phi) \in C^{\infty}\Biggl(\biggl(- \frac{\pi}{2}, \frac{\pi}{2}\biggr) \times [0, 2 \pi), \Biggl(0, \sqrt{\frac{2 q}{\smash{1 + \sin{(\vartheta)}}}} \,\, \Biggr]\Biggr) 
\end{equation*}
determines the geometrical shape of the termination shock, we obtain a lower bound for the radial image
\begin{equation*} 
\textnormal{Ran}(r) = \Biggl[R(\vartheta, \phi), \sqrt{\frac{2 q}{\smash{1 + \sin{(\vartheta)}}}} \,\, \Biggr] \, .
\end{equation*}

\begin{figure}[t]%
\centering
\includegraphics[width=0.60\columnwidth]{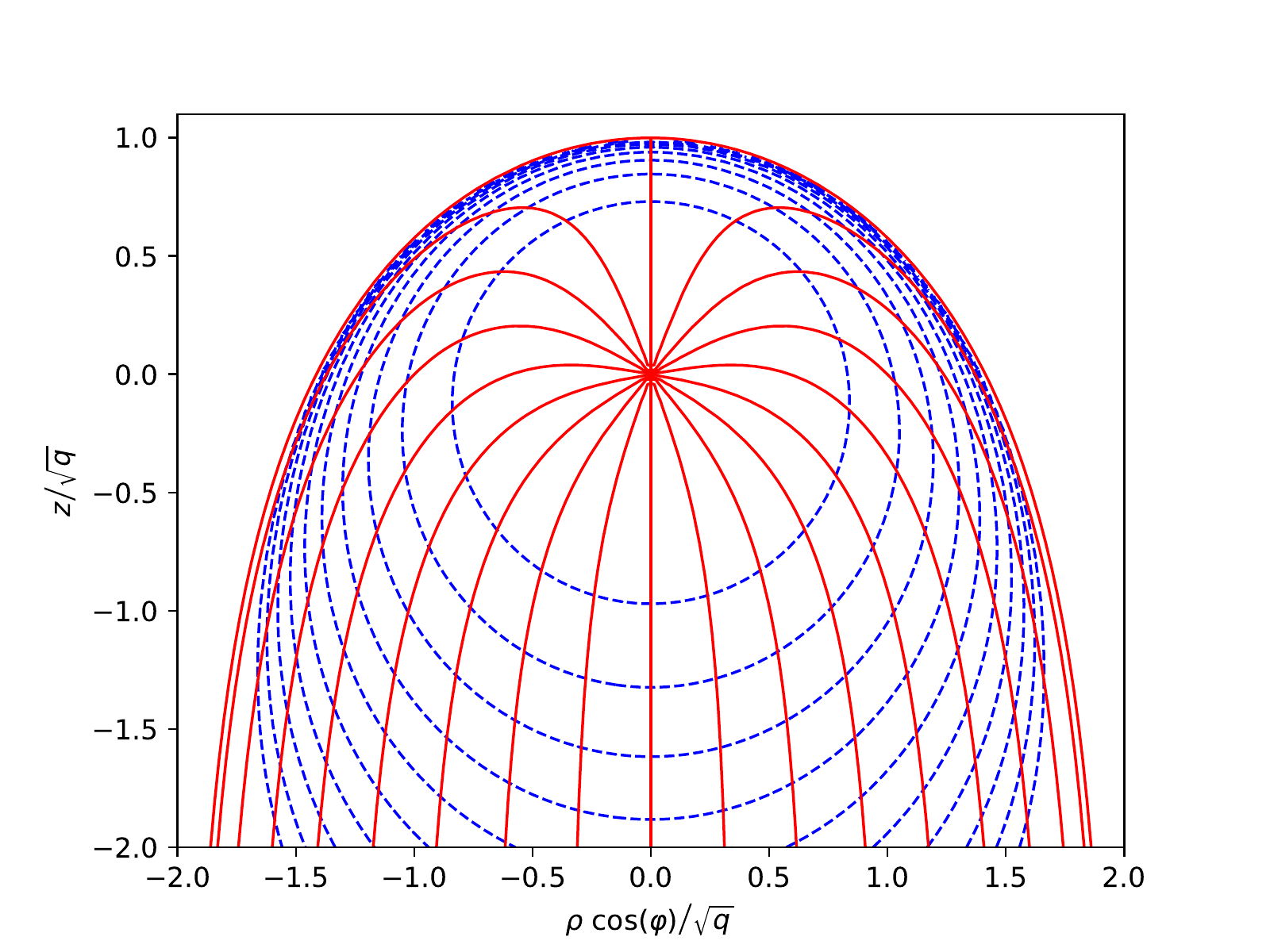}%
\caption[...]
{Contours of the coordinates $w \in (- q, q)$ (red, solid curves) and $\mu \in [- 2 q^{3/2}, 0]$ (blue, dashed curves) for $T_0 = 0$ in an arbitrarily chosen plane of constant $\varphi$, using equidistant spacings for $\arcsin{(w)}$ and $\mu$ to ensure that all streamlines meet in the origin with uniform $10^{\circ}$ angular separation. The $\mu = 0$ contour is not visible as it is a single point located at the origin.}%
\label{helio4_psi+phi}%
\end{figure}

Next, for the purpose of reducing the ideal Ohm's law (\ref{OhmSimp}) to a system of ordinary differential equations, viz., to have a representation in characteristic coordinates, we perform two consecutive coordinate transformations. The first of these transformations is given by 
\begin{equation*} 
\mathfrak{T}^{(3)} \hspace{-0.05cm}:
\begin{cases}
\, \displaystyle I \times \Biggl[R(\vartheta, \phi), \sqrt{\frac{2 q}{\smash{1 + \sin{(\vartheta)}}}} \,\, \Biggr] \times \biggl(- \frac{\pi}{2}, \frac{\pi}{2}\biggr) \times [0, 2 \pi) \rightarrow I' \times \biggl(- \frac{\pi}{2}, \frac{\pi}{2}\biggr) \times \bigl[v_{\textnormal{B}}(u, \Phi), q\bigr) \times [0, 2 \pi) \\ \\
\hspace{6.40cm} (t, r, \vartheta, \phi) \mapsto (T, u, v, \Phi)
\end{cases} 
\end{equation*}
with
\begin{equation} \label{T2C}
T = u_0 q \, t \, , \quad u = \vartheta \, , \quad v = \frac{r^2 \cos^2{(\vartheta)}}{2} + q \sin{(\vartheta)} \, , \quad \textnormal{and} \quad \Phi = \phi \, ,
\end{equation}
where 
\begin{equation*}
I' := u_0 q \, I \quad \textnormal{and} \quad v_{\textnormal{B}}(u, \Phi) := \frac{R^2(u, \Phi) \cos^2{(u)}}{2} + q \sin{(u)} \, .
\end{equation*}
The second transformation is of the form
\begin{equation} \label{T3}
\mathfrak{T}^{(4)} \hspace{-0.05cm}:
\begin{cases}
\, \displaystyle I' \times \biggl(- \frac{\pi}{2}, \frac{\pi}{2}\biggr) \times \bigl[v_{\textnormal{B}}(u, \Phi), q\bigr) \times [0, 2 \pi) \rightarrow I' \times \bigl(- \infty, \textnormal{max}(I')\bigr) \times \bigl[w_{\textnormal{B}}(\mu - \tau, \chi), q\bigr) \times [0, 2 \pi) \\ \\
\hspace{4.41cm} (T, u, v, \Phi) \mapsto (\tau, \mu, w, \chi)
\end{cases} 
\end{equation}
with
\begin{equation} \label{T3C}
\tau = T \, , \quad \mu = T + q \, \mathscr{F}(u, v) \, , \quad w = v \, , \quad \textnormal{and} \quad \chi = \Phi \, ,
\end{equation}
in which  
\begin{equation} \label{F}
\mathscr{F}(u, v) := \int \frac{r(u, v)}{\cos{(u)}} \, \textnormal{d}u
\end{equation}
and $w_{\textnormal{B}}(\mu - \tau, \chi)$ is a solution to the equation 
\begin{equation} \label{wBound}
R\bigl(\mathscr{F}^{- 1}_{u}([\mu -\tau]/q, w), \chi\bigr) = \frac{\sqrt{2 \, \bigl[w - q \sin{\bigl(\mathscr{F}^{- 1}_{u}([\mu -\tau]/q, w)\bigr)}\bigr]}}{\cos{\bigl(\mathscr{F}^{- 1}_{u}([\mu -\tau]/q, w)\bigr)}} 
\end{equation}
with respect to the coordinate $w$. An explicit analytical expression for the function $\mathscr{F}$ by means of incomplete elliptic integrals of the first and second kind can be found in Appendix A. Also, the quantity $\mathscr{F}^{- 1}_{u}(\, . \, , . \, )$ denotes the inverse of this function with respect to the variable $u$ at $v = w$. We point out that, since $\mathscr{F}$ is a continuously differentiable function and $\partial_u \mathscr{F}$ is invertible, the existence of such an inverse results directly from the implicit function theorem. Moreover, the coordinate $w$ may be interpreted as a label for streamlines emanating from their common solar origin, while hypersurfaces of constant $\mu$ form isochrones, which connect fluid elements that were emitted at the origin at a common time $T = T_0$ (see FIG.\ \ref{helio4_psi+phi}).

\subsection{General Magnetic Vector Potential and Magnetic Field Solutions} \label{SecIIIB}

\noindent For the derivation of a general magnetic field solution for the inner heliosheath, one may employ the limit (\ref{induction}) of the induction equation in combination with the flow field (\ref{flowfield}), and impose the magnetic divergence constraint (\ref{mdc}). In terms of the characteristic coordinates defined in (\ref{T3C}), this yields, i.a., the Schr\"odinger-type equations
\begin{equation*}
\bigl[\partial_{\tau \tau} + V_k(\mu - \tau, w)\bigr] \bigl(\mathscr{W}_k(\mu - \tau, w) \, B_k\bigr) = 0 \, ,
\end{equation*}
where the potentials $V_k \hspace{-0.07cm} : I \times \overline{\Omega} \rightarrow \mathbb{R}$ are smooth and decay quadratically for large distances, $\mathscr{W}_k \hspace{-0.07cm} : I \times \overline{\Omega} \rightarrow \mathbb{R}_{> 0}$ denote weight functions, and $k \in \{\rho, z\}$. As the explicit functional shapes of the potentials are rather intricate, these Schr\"odinger-type equations are quite difficult to solve. We avoid this problem by considering the ideal Ohm's law (\ref{OhmSimp}) and the magnetic curl relation (\ref{BArel}) instead. Thus, working with the cylindrical coordinates (\ref{T0C}) and again prescribing the flow field (\ref{flowfield}), the ideal Ohm's law reads 
\begin{align}
u_0 \biggl(1 - \frac{q z}{(\rho^2 + z^2)^{3/2}}\biggr) [\partial_z A_{\rho} - \partial_{\rho} A_z] & = \partial_t A_{\rho} + \partial_{\rho} \psi \label{e1} \\ \nonumber \\
\frac{u_0 q \rho}{(\rho^2 + z^2)^{3/2}} \, [\partial_z A_{\rho} - \partial_{\rho} A_z] & = \partial_t A_z + \partial_z \psi \label{e2} \\ \nonumber \\
\frac{u_0 q}{(\rho^2 + z^2)^{3/2}} \biggl[\partial_{\varphi} A_{\rho} - \partial_{\rho}(\rho A_{\varphi}) + \biggl(z - \frac{(\rho^2 + z^2)^{3/2}}{q}\biggr) \biggl(\frac{1}{\rho} \, \partial_{\varphi} A_z - \partial_z A_{\varphi}\biggr)\biggr] & = \partial_t A_{\varphi} + \frac{1}{\rho} \, \partial_{\varphi} \psi \, . \label{e3}
\end{align}
Combining equations (\ref{e1}) and (\ref{e2}) to 
\begin{equation} \label{combi12}
\frac{u_0 q}{(\rho^2 + z^2)^{3/2}} \biggl[\rho \, \partial_{\rho} + \biggl(z - \frac{(\rho^2 + z^2)^{3/2}}{q}\biggr) \partial_z\biggr] \psi = - \partial_t (\boldsymbol{u} \cdot \boldsymbol{A}) 
\end{equation}
by eliminating the term $\partial_z A_{\rho} - \partial_{\rho} A_z$, we can determine the electric scalar potential $\psi$ by using a suitable gauge condition. To be more precise, expressing equation (\ref{combi12}) via the characteristic coordinates (\ref{T3C}), we obtain the representation 
\begin{equation*} 
\partial_{\tau} \psi = [\partial_{\tau} + \partial_{\mu}] (\psi - \boldsymbol{u} \cdot \boldsymbol{A}) \, .
\end{equation*}
From this representation, it is obvious to choose the gauge condition 
\begin{equation} \label{agc}
\psi = \boldsymbol{u} \cdot \boldsymbol{A} \, ,
\end{equation}
which leads to the ordinary differential equation 
\begin{equation*}
\partial_{\tau} \psi = 0 \, .
\end{equation*}
Hence, the solution for the electric scalar potential is the constant of integration 
\begin{equation} \label{solesp}
\psi = \psi(\mu, w, \chi) \in C^2(I \times \Omega, \mathbb{R}) \, ,
\end{equation}
whose explicit functional shape depends on the particular choice of the initial-boundary values at $\{\tau_{\textnormal{B}}\} \times \partial \Omega$. We note that the gauge condition (\ref{agc}) is an axial gauge with a residual gauge freedom. Therefore, it only accounts for a partial gauge fixing. Next, writing the gauge condition in the form
\begin{equation} \label{trgc}
A_{\rho} = \frac{(\rho^2 + z^2)^{3/2}}{u_0 q \rho} \biggl[\psi + u_0 \biggl(1 - \frac{q z}{(\rho^2 + z^2)^{3/2}}\biggr) A_z\biggr] \, ,
\end{equation}
and inserting this expression --- as well as the solution for the scalar potential (\ref{solesp}) --- into equation (\ref{e2}), we find the inhomogeneous first-order partial differential equation 
\begin{equation*}  
\biggl[\frac{(\rho^2 + z^2)^{3/2}}{u_0 q} \, \partial_t + \rho \, \partial_{\rho} + \biggl(z - \frac{(\rho^2 + z^2)^{3/2}}{q}\biggr) \partial_z - \frac{3 \sqrt{\rho^2 + z^2} \, z}{q} + 1\biggr] A_z = \frac{3 \sqrt{\rho^2 + z^2} \, z \psi}{u_0 q} \, .
\end{equation*}
In order to eliminate the zero-order terms, we make the ansatz
\begin{equation} \label{anszc}
A_z(t, \rho, z, \varphi) = \frac{\rho^2}{(\rho^2 + z^2)^{3/2}} \, \mathscr{C}_z(t, \rho, z, \varphi) \, ,
\end{equation}
which yields the equation
\begin{equation*}
\biggl[\frac{(\rho^2 + z^2)^{3/2}}{u_0 q} \, \partial_t + \rho \, \partial_{\rho} + \biggl(z - \frac{(\rho^2 + z^2)^{3/2}}{q}\biggr) \partial_z\biggr] \mathscr{C}_z = \frac{3 \, (\rho^2 + z^2)^{2} \, z \psi}{u_0 q \rho^2} \, .
\end{equation*}
Using once again the characteristic coordinates (\ref{T3C}), we obtain the ordinary differential equation
\begin{equation*}
\partial_{\tau} \mathscr{C}_z = \frac{3 r z \psi}{u_0 q \rho^2} \, .
\end{equation*}
Integration gives rise to the solution
\begin{equation*} 
\mathscr{C}_z(\tau, \mu, w, \chi) = \frac{3 \psi(\mu, w, \chi) \, \mathcal{J}(\mu - \tau, w)}{u_0 q} + \mathcal{H}(\mu, w, \chi) \in C^2(I \times \Omega, \mathbb{R}) \, ,
\end{equation*}
where 
\begin{equation} \label{FJ} 
\mathcal{J}(\mu - \tau, w) := \int \frac{r z}{\rho^2} \, \textnormal{d}\tau
\end{equation}
and $\mathcal{H}(\mu, w, \chi)$ is a constant of integration. We derive an explicit analytical representation of the function $\mathcal{J}$ in terms of incomplete elliptic integrals of the first and second kind in Appendix A. Accordingly, the $z$-component (\ref{anszc}) of the magnetic vector potential becomes 
\begin{equation} \label{sol1}
A_z = \frac{\rho^2}{r^3} \biggl[\frac{3 \psi \mathcal{J}}{u_0 q} + \mathcal{H}\biggr] \, . 
\end{equation}
Moreover, substituting this solution and the scalar potential (\ref{solesp}) into the gauge condition (\ref{trgc}), we immediately find the $\rho$-component 
\begin{equation} \label{sol2}
A_{\rho} = \frac{r^3 \psi}{u_0 q \rho} - \rho \biggl(\frac{z}{r^3} - \frac{1}{q}\biggr) \biggl[\frac{3 \psi \mathcal{J}}{u_0 q} + \mathcal{H}\biggr] \, .
\end{equation}
For the determination of the azimuthal component, we insert the gauge condition (\ref{agc}) as well as the solutions (\ref{sol1}) and (\ref{sol2}) into equation (\ref{e3}), which results in the homogeneous first-order partial differential equation
\begin{equation*} 
\biggl[\frac{(\rho^2 + z^2)^{3/2}}{u_0 q} \, \partial_t + \rho \, \partial_{\rho} + \biggl(z - \frac{(\rho^2 + z^2)^{3/2}}{q}\biggr) \partial_z + 1\biggr] A_{\varphi} = 0 \, .
\end{equation*}
The zero-order term can be eliminated via the ansatz
\begin{equation} \label{sol3}
A_{\varphi}(t, \rho, z, \varphi) = \frac{1}{\rho} \, \mathscr{C}_{\varphi}(t, \rho, z, \varphi) \, ,
\end{equation}
leading to the equation
\begin{equation*}
\biggl[\frac{(\rho^2 + z^2)^{3/2}}{u_0 q} \, \partial_t + \rho \, \partial_{\rho} + \biggl(z - \frac{(\rho^2 + z^2)^{3/2}}{q}\biggr) \partial_z\biggr] \mathscr{C}_{\varphi} = 0 \, .
\end{equation*}
By means of the characteristic coordinates (\ref{T3C}), this equation reads
\begin{equation*}
\partial_{\tau} \mathscr{C}_{\varphi} = 0 \, .
\end{equation*}
Integration yields
\begin{equation} \label{Cphi}
\mathscr{C}_{\varphi} = \mathscr{C}_{\varphi}(\mu, w, \chi) \in C^2(I \times \Omega, \mathbb{R}) \, .
\end{equation}
Finally, we substitute the solutions (\ref{sol1}), (\ref{sol2}), and (\ref{sol3}) together with (\ref{Cphi}) into the magnetic curl relation (\ref{BArel}), and express the cylindrical partial derivative operators in the characteristic form
\begin{equation*}
\partial_{\rho} = \rho \biggl(1 - \frac{q z}{r^3}\biggr) \partial_{w} - \biggl[\frac{r^3}{\rho} + \frac{3 \rho \mathcal{J}}{q} \biggl(1 - \frac{q z}{r^3}\biggr)\biggr] \partial_{\mu} \, , \quad \partial_z = \frac{\rho^2}{r^3} \bigl[q \, \partial_{w} - 3 \mathcal{J} \partial_{\mu}\bigr] \, , \quad \textnormal{and} \quad \partial_{\varphi} = \partial_{\chi} \, .
\end{equation*}
This gives rise to the magnetic field representation
\begin{subequations} \label{Bcomp} 
\begin{align}	
B_{\rho} & = \frac{\rho}{r^3} \biggl[3 \mathcal{J} \biggl(\frac{\partial_{\chi} \psi}{u_0 q} + \partial_{\mu} \mathscr{C}_{\varphi}\biggr) + \partial_{\chi} \mathcal{H} - q \, \partial_w \mathscr{C}_{\varphi}\biggr] \\ \nonumber \\
B_z & = \biggl(\frac{z}{r^3} - \frac{1}{q}\biggr) \biggl[3 \mathcal{J} \biggl(\frac{\partial_{\chi} \psi}{u_0 q} + \partial_{\mu} \mathscr{C}_{\varphi}\biggr) + \partial_{\chi} \mathcal{H} - q \, \partial_w \mathscr{C}_{\varphi}\biggr] - \frac{r^3}{\rho^2} \biggl(\frac{\partial_{\chi} \psi}{u_0 q} + \partial_{\mu} \mathscr{C}_{\varphi}\biggr) \\ \nonumber \\
B_{\varphi} & = \rho \biggl(\frac{\partial_w \psi}{u_0} + \partial_{\mu} \mathcal{H}\biggr) \, ,
\end{align}
\end{subequations}
which is the principal result of the paper. It should be pointed out that the above method may be applied to any time-independent, curl-free flow field exhibiting at least one spatial symmetry. Therefore, it also works for more general flow fields with arbitrary divergence.

\section{Initial-Boundary Conditions} \label{SecIV}

\subsection{Well-posedness of the Initial-Boundary Value Problem and Compatibility of Initial-Boundary Values} \label{SecIVA}

\noindent Thus far, we have derived a general solution $\boldsymbol{B} \in C^1(I \times \Omega, \mathbb{R}^3)$ of both the limit (\ref{redineq}) of the induction equation for a prescribed flow field of the form (\ref{flowfield}) and the magnetic divergence constraint (\ref{mdc}). To obtain a particular solution in the class $C^1(I \times \Omega, \mathbb{R}^3) \cap C^0(I \times \overline{\Omega}, \mathbb{R}^3)$, one has to solve the associated Dirichlet-type initial-boundary value problem that also comprises the initial-boundary condition (\ref{BV}). Here, we evaluate the well-posedness of this initial-boundary value problem applying the notion introduced by Hadamard \cite{Had}, which consists of the following three conditions:
\begin{itemize}
\item[($\mathcal{C}1$)] There exists a solution to the initial-boundary value problem.
\item[($\mathcal{C}2$)] The solution is unique.
\item[($\mathcal{C}3$)] The solution is stable, that is, it depends continuously on the initial-boundary values.
\end{itemize}
In order to verify the first two Hadamard conditions, we employ results from the method of characteristics for first-order systems of partial differential equations (see, e.g., \cite{evans}). We start with the determination of an explicit parametric representation of the projected characteristics by solving the system of Lagrange--Charpit equations
\begin{equation} \label{LCE}
\frac{\textnormal{d}t}{\textnormal{d}\tau} = \frac{1}{u_0 q} \, , \quad \frac{\textnormal{d}r}{\textnormal{d}\tau} = \frac{1}{r^2} - \frac{\sin{(\vartheta)}}{q} \, , \quad \frac{\textnormal{d}\vartheta}{\textnormal{d}\tau} = - \frac{\cos{(\vartheta)}}{r q} \, , \quad \textnormal{and} \quad \frac{\textnormal{d}\phi}{\textnormal{d}\tau} = 0 \, 
\end{equation}
in the spherical coordinate representation defined by (\ref{SphericalCR}), where the time parameter $\tau \in I'$ specified in (\ref{T3}) is used for the parametrization. The projected characteristics thus yield
\begin{equation*} 
\boldsymbol{\mathcal{C}}_{\tau} \hspace{-0.05cm}:
\begin{cases}
\, I' \rightarrow I \times \overline{\Omega} \vspace{0.2cm} \\ 
\hspace{0.13cm} \tau \mapsto \bigl(t(\tau), r(\tau), \vartheta(\tau), \phi(\tau)\bigr)  
\end{cases} 
\end{equation*}
with 
\begin{equation} \label{projchar} 
t(\tau) = \frac{\tau}{u_0 q} + \alpha_0 \, , \quad r(\tau) = \frac{\sqrt{2 \, \bigl[\alpha_1 - q \sin{\bigl(\vartheta(\tau)\bigr)}\bigr]}}{\cos{\bigl(\vartheta(\tau)\bigr)}} \, , \quad \vartheta(\tau) = \mathscr{F}^{- 1}_{u}(-\tau/q + \alpha_2, \alpha_1) \, , \quad \textnormal{and} \quad \phi(\tau) = \alpha_3 \, ,
\end{equation}
in which the quantities $\alpha_l$, $l \in \{0, 1, 2, 3\}$, are constants of integration. We remark that the characteristic coordinates (\ref{T3C}) correspond to the particular choice $\alpha_0 = 0$, $\alpha_1 = w$, $\alpha_2 = \mu/q$, and $\alpha_3 = \chi$. Next, we solve the characteristic equation for the inner boundary surface $I \times \partial \Omega$
\begin{equation} \label{CE}
\bigg\langle\frac{\textnormal{d}\boldsymbol{\mathcal{C}}_{\tau}}{\textnormal{d}\tau} \, , \, \boldsymbol{n}\bigg\rangle_{| I \times \partial \Omega \, \cap \, \boldsymbol{\mathcal{C}}_{\tau}} = 0 \, ,
\end{equation}
where $\langle \, . \, , . \, \rangle = \boldsymbol{\eta}(\, . \, , . \,)$ is the standard scalar product on $I \times \overline{\Omega}$ and $\boldsymbol{n}$ is an outward-pointing normal to $I \times \partial \Omega$. To compute the normal, we first express the inner boundary in terms of the implicit surface equation (\ref{boundarypar}), and then evaluate the orthogonality conditions
\begin{equation*}
\langle \boldsymbol{n}, \boldsymbol{e}_t \rangle_{|I \times \partial \Omega} = 0 \, , \quad \langle \boldsymbol{n}, \partial_{\vartheta}R \, \boldsymbol{e}_r + R \, \boldsymbol{e}_{\vartheta} \rangle_{|I \times \partial \Omega} = 0 \, , \quad \textnormal{and} \quad \langle \boldsymbol{n}, \partial_{\phi}R \, \boldsymbol{e}_r + R \sin{(\vartheta)} \, \boldsymbol{e}_{\phi} \rangle_{|I \times \partial \Omega} = 0 \, . 
\end{equation*}
Inserting the result and the Lagrange--Charpit equations (\ref{LCE}) into (\ref{CE}), we obtain 
\begin{equation*}
R(\vartheta, \phi) = \frac{\sqrt{\mathcal{G}(\phi) - 2 q \sin{(\vartheta)}}}{\cos{(\vartheta)}} \quad \,\, \textnormal{with} \quad \mathcal{G}(\phi) \in C^{\infty}\bigl([0, 2 \pi), [-q, 2 q)\bigr)  
\end{equation*}
for some neighborhood $U \subset I \times \partial \Omega$ of $I \times \partial \Omega \, \cap \, \boldsymbol{\mathcal{C}}_{\tau}$. Therefore, $I \times \partial \Omega$ is noncharacteristic --- and hence transverse to all projected characteristics --- if and only if $R(\vartheta, \phi)$ is not of this particular form for any neighborhood $U$ in $I \times \partial \Omega$. Moreover, employing the expressions in (\ref{projchar}) with the subjacent choice for the constants of integration $\alpha_l$, the intersection condition $\boldsymbol{\mathcal{C}}_{\tau} = (t, \boldsymbol{x})_{| I \times \partial \Omega \, \cap \, \boldsymbol{\mathcal{C}}_{\tau}}$ yields equation (\ref{wBound}) evaluated at $I \times \partial \Omega \, \cap \, \boldsymbol{\mathcal{C}}_{\tau}$. Thus, in case $R(\vartheta, \phi)$ is further chosen such that this equation has exactly one solution $\tau \in I'$ for every $\boldsymbol{\mathcal{C}}_{\tau}$, the inner boundary surface intersects each projected characteristic once only. Also, as the computation of the Jacobian determinant results in
\begin{equation*}
\textnormal{det}(\boldsymbol{J}) = \frac{\partial(t, r, \vartheta, \phi)}{\partial(\tau, \mu, w, \chi)} = - \frac{1}{u_0 q^2 r^2 \cos{(\vartheta)}} \notin \{0, \pm \infty\} \, , 
\end{equation*}
the family of projected characteristics $(\boldsymbol{\mathcal{C}}_{\tau})_{\tau \in I'}$ is immediately proven to be space-filling and nonintersecting. Taken together, this guarantees the existence of a unique, global solution to the initial-boundary value problem. The validity of the third Hadamard condition, however, depends on the particular choice of the initial-boundary values, and thus this condition has to be evaluated case by case. 

Finally, we note that in order for the Dirichlet-type initial-boundary values $\boldsymbol{g}$ to be admissible in the first place, they have to satisfy a consistency condition as well as a necessary condition. These are stated in the following. Firstly, the initial-boundary values are required to vary along the inner boundary surface $\{t_{\textnormal{B}}\} \times \partial \Omega$ in the same way as the magnetic field according to both the limit (\ref{redineq}) of the induction equation and the magnetic divergence constraint (\ref{mdc}). As a consequence, initial-boundary values have to be chosen such that they are compatible with the specific form of the magnetic field solution (\ref{Bcomp}) evaluated at $\{t_{\textnormal{B}}\} \times \partial \Omega$. Secondly, integrating the magnetic divergence constraint over the domain $\overline{\Omega}$ and using Stokes' theorem, we obtain the additional condition
\begin{equation*} 
\int_{\partial \Omega} \boldsymbol{g} \cdot \boldsymbol{\mathfrak{n}} \, \textnormal{d}\mu_{\partial \Omega} = 0 \, ,
\end{equation*}
where $\boldsymbol{\mathfrak{n}}$ is the outward-pointing unit normal to --- and $\textnormal{d}\mu_{\partial \Omega}$ is the invariant measure on --- $\partial \Omega$.

\subsection{Implementation of Initial-Boundary Conditions} \label{SecIVB}

\noindent We now outline the general procedure of imposing suitable initial-boundary values on our general magnetic field solution (\ref{Bcomp}). To this end, we consider Dirichlet-type initial-boundary values $\boldsymbol{g} \in C^0\bigl(\{t_{\textnormal{B}}\} \times \partial \Omega, \mathbb{R}^3\bigr)$ that are, on the one hand, admissible in the above sense and, on the other hand, yield a direct link to the structure of the interplanetary magnetic field. As the latter magnetic field is usually expressed via the spherical coordinates (\ref{ASC}) defined in Appendix B, we have to write our magnetic field solution in terms of these coordinates (see formula (\ref{mfc})). Evaluating this representation at $\{t_{\textnormal{B}}\} \times \partial \Omega$ and equating the result with the prescribed Dirichlet-type initial-boundary values, we obtain a system of linear algebraic equations for the functions $\mathcal{T}_i$, $i \in \{1, 2, 3\}$, which are defined below formula (\ref{mfc}), on the inner boundary. As the solution of this system is constant along each projected characteristic, it can be analytically extended to the domain $I \times \overline{\Omega}$. This gives rise to a particular magnetic field solution for the inner heliosheath. We note in passing that in order to see that this particular solution indeed satisfies the magnetic divergence constraint, one can simply verify the validity of the relation 
\begin{equation*} 
q \, \partial_w \mathcal{T}_1 + \partial_{\mu} \mathcal{T}_2 = \partial_{\chi} \mathcal{T}_3 \, .
\end{equation*}

\section{Summary and Outlook} \label{SecV}

\noindent In the present paper, we have continued our analytical considerations of the large-scale magnetic field structure of the heliosphere and its outer vicinity by transferring our earlier modeling to the inner heliosheath. Since the structure of this region is inherently time-dependent as a consequence of solar activity variations in the supersonic solar wind inside the termination shock, we have thus extended our modeling to time-dependence. More precisely, we have derived an exact analytical expression for the time-dependent magnetic field in the inner heliosheath by first solving a characteristic coordinate representation of the ideal Ohm's law for the magnetic vector potential prescribing a Rankine-type flow field and choosing an axial gauge, and then making use of the magnetic curl relation. We also showed that this magnetic field expression is a general solution of both the induction equation of ideal magnetohydrodynamics in the limit of infinite electric conductivity as well as the magnetic divergence constraint. In contrast to finding such a general solution, it is of course far more intricate to obtain a particular solution for modeling observational data, as this involves solving a Dirichlet-type initial-boundary value problem for which a careful specification of the boundary geometry and the initial-boundary conditions is required. Nonetheless, we will fulfill this task in comprehensive detail in a separate paper, where we will furthermore work out an alternative method of derivation for the magnetic field in the inner heliosheath by means of the Cauchy integral formalism.

\section*{Acknowledgments}
\noindent C.R.\ is grateful to Katharina Proksch and Florian Schuppan for useful discussions and comments. C.R. is partially supported by the research project MTM2016-78807-C2-1-P funded by MINECO and ERDF. J.K.\ and H.F.\ are grateful for financial support from the DFG within the framework of the research project FI 706/23-1.

\begin{appendix}

\section*{Appendix A: Elliptic Integral Representations of $\mathscr{F}$ and $\mathcal{J}$} \label{AppA}

\noindent We determine elliptic integral representations of the functions $\mathscr{F}$ and $\mathcal{J}$ defined in (\ref{F}) and (\ref{FJ}), respectively. Beginning with the former, we substitute the function $r(u, v) = \sqrt{2 \, [v - q \sin{(u)}]}/\cos{(u)}$ according to (\ref{T2C}) and apply integration by parts, which results in
\begin{equation} \label{trint}
\frac{\mathscr{F}(u, v)}{\sqrt{2}} = \sqrt{v - q \sin{(u)}} \, \tan{(u)} + \frac{q}{2} \int \frac{\sin{(u)}}{\sqrt{v - q \sin{(u)}}} \, \textnormal{d}u \, .
\end{equation}
To solve the residual integral, it is advantageous to first perform the splitting 
\begin{equation} \label{expr2} 
\int \frac{\sin{(u)}}{\sqrt{v - q \sin{(u)}}} \, \textnormal{d}u = \frac{v}{q} \int \frac{1}{\sqrt{v - q \sin{(u)}}} \, \textnormal{d}u - \frac{1}{q} \int \sqrt{v - q \sin{(u)}} \, \textnormal{d}u \, .
\end{equation}
Then, using the coordinate transformation
\begin{equation} \label{inttrafo}
\mathfrak{T}^{(5)} \hspace{-0.05cm}:
\begin{cases}
\, \displaystyle \biggl(- \frac{\pi}{2}, \frac{\pi}{2}\biggr) \rightarrow (0, 1) \vspace{0.2cm} \\ 
\hspace{1.39cm} \displaystyle u \mapsto m = \sqrt{\frac{v - q \sin{(u)}}{v + q}} \, ,
\end{cases} 
\end{equation}
we obtain
\begin{equation} \label{Simpint}
\int \frac{1}{\sqrt{v - q \sin{(u)}}} \, \textnormal{d}u = - \frac{2}{\sqrt{q - v}} \int \frac{1}{\sqrt{(1 - m^2) (1 + \lambda m^2)}} \, \textnormal{d}m 
\end{equation}
and
\begin{equation} \label{Simpint2}
\int \sqrt{v - q \sin{(u)}} \, \textnormal{d}u = 2 \sqrt{q - v} \, \Biggl[\int \frac{1}{\sqrt{(1 - m^2) (1 + \lambda m^2)}} \, \textnormal{d}m - \int \sqrt{\frac{1 + \lambda m^2}{1 - m^2}} \, \textnormal{d}m\Biggr] \, ,
\end{equation}
where $\lambda := (q + v)/(q - v)$. The right hand sides of these equations can be directly identified as incomplete elliptic integrals of the first and second kind in Legendre normal form
\begin{equation*}
F(m, n) := \int_0^m \frac{1}{\sqrt{(1 - m'^2) (1 - n^2 m'^2)}} \, \textnormal{d}m' \quad \textnormal{and} \quad E(m, n) := \int_0^m \sqrt{\frac{1 - n^2 m'^2}{1 - m'^2}} \, \textnormal{d}m'
\end{equation*}
for $n = \textnormal{i} \sqrt{\lambda}$. Now, inserting (\ref{Simpint}) and (\ref{Simpint2}) into (\ref{expr2}) yields
\begin{equation} \label{split2}
\int \frac{\sin{(u)}}{\sqrt{v - q \sin{(u)}}} \, \textnormal{d}u = 2 \, \biggl[\frac{\sqrt{q - v}}{q} \, E\bigl(m, \textnormal{i} \sqrt{\lambda} \, \bigr) - \frac{1}{\sqrt{q - v}} \, F\bigl(m, \textnormal{i} \sqrt{\lambda} \, \bigr)\biggr] \, .
\end{equation}
Since we require the second arguments of the elliptic integrals to be real-valued, we employ the identities
\begin{equation} \label{ID1} 
F\bigl(m, \textnormal{i} \sqrt{\lambda} \, \bigr) = \frac{1}{\sqrt{\lambda}} \, F\Biggl(\sqrt{\frac{\lambda m^2}{1 + \lambda m^2}}, \, \sqrt{1 + \frac{1}{\lambda}} \,\, \Biggr) 
\end{equation}
and
\begin{equation} \label{ID2} 
E\bigl(m, \textnormal{i} \sqrt{\lambda} \, \bigr) = \frac{1}{\sqrt{\lambda}} \, F\Biggl(\sqrt{\frac{\lambda m^2}{1 + \lambda m^2}}, \, \sqrt{1 + \frac{1}{\lambda}} \,\, \Biggr) + \sqrt{\lambda} \, E\Biggl(\sqrt{\frac{\lambda m^2}{1 + \lambda m^2}}, \, \sqrt{1 + \frac{1}{\lambda}} \,\, \Biggr) - \lambda m \, \sqrt{\frac{1 - m^2}{1 + \lambda m^2}} \, .
\end{equation}
Substituting the resulting representation of the integral (\ref{split2}) into (\ref{trint}), we find
\begin{equation*} 
\begin{split}
\frac{\mathscr{F}(u, v)}{\sqrt{2}} & = \sqrt{v - q \sin{(u)}} \, \Biggl[\tan{(u)} - \sqrt{\frac{1 + \sin{(u)}}{1 - \sin{(u)}}} \,\, \Biggr] + \sqrt{v + q} \, E\Biggl(\sqrt{\frac{v - q \sin{(u)}}{q \, [1 - \sin{(u)}]}}, \, \sqrt{\frac{2 q}{v + q}} \, \Biggr) \\ \\
& \hspace{0.4cm}- \frac{v}{\sqrt{v + q}} \, F\Biggl(\sqrt{\frac{v - q \sin{(u)}}{q \, [1 - \sin{(u)}]}}, \, \sqrt{\frac{2 q}{v + q}} \, \Biggr) \, .
\end{split}
\end{equation*}
In terms of the cylindrical coordinates (\ref{T0C}) and with $r = r(\rho, z) = \sqrt{\rho^2 + z^2}$, this expression reads
\begin{equation*}
\begin{split}
\mathscr{F}(\rho, z) & = - r + \sqrt{\rho^2 + 2 q \, (1 + z/r)} \, E\Biggl(\frac{\rho}{\sqrt{2 q \, (1 - z/r)}}, \, 2 \, \sqrt{\frac{q}{\rho^2 + 2 q \, (1 + z/r)}} \, \Biggr) \\ \\
& \hspace{0.4cm}- \frac{\rho^2 + 2 q z/r}{\sqrt{\rho^2 + 2 q \, (1 + z/r)}} \, F\Biggl(\frac{\rho}{\sqrt{2 q \, (1 - z/r)}}, \, 2 \, \sqrt{\frac{q}{\rho^2 + 2 q \, (1 + z/r)}} \, \Biggr) \, .
\end{split}
\end{equation*}
The elliptic integral representation of the function $\mathcal{J}$ may be derived in a similar vein as shown below. We start by writing (\ref{FJ}) in the form
 \begin{equation*}
\mathcal{J} = - q \int \frac{r(u, v) \sin{(u)}}{\cos^3(u)} \, \textnormal{d}u \, .
\end{equation*}
Also inserting the above function $r(u, v)$ and applying two-fold integration by parts, we obtain
\begin{equation} \label{I21}
\frac{\mathcal{J}}{q \, \sqrt{2}} = - \frac{\sqrt{v - q \sin{(u)}}}{3 \cos^3(u)} - \frac{q \tan{(u)}}{6 \sqrt{v - q \sin{(u)}}} + \frac{q^2}{12} \int \frac{\sin{(u)}}{[v - q \sin{(u)}]^{3/2}} \, \textnormal{d}u \, .
\end{equation}
The residual integral can be split into the two integrals
\begin{equation} \label{s32int}
\int \frac{\sin{(u)}}{[v - q \sin{(u)}]^{3/2}} \, \textnormal{d}u = \frac{v}{q} \int \frac{1}{[v - q \sin{(u)}]^{3/2}} \, \textnormal{d}u - \frac{1}{q} \int \frac{1}{\sqrt{v - q \sin{(u)}}} \, \textnormal{d}u \, ,
\end{equation}
where the second integral on the right hand side is already given by formula (\ref{Simpint}). To calculate the first integral, we again use the transformation (\ref{inttrafo}) as well as another splitting, which leads to the following representation in terms of incomplete elliptic integrals of the first and second kind
\begin{equation} \label{32int}
\begin{split}
\int \frac{1}{[v - q \sin{(u)}]^{3/2}} \, \textnormal{d}u & = - \frac{2}{\sqrt{q - v} \, (q + v)} \int \frac{1 + \lambda m^4 - \lambda m^4}{m^2 \, \sqrt{(1 - m^2) (1 + \lambda m^2)}} \, \textnormal{d}m \\ \\
& = \frac{2}{\sqrt{q - v} \, (q + v)} \Biggl[\int \frac{\textnormal{d}}{\textnormal{d}m} \biggl(\frac{\sqrt{(1 - m^2) (1 + \lambda m^2)}}{m} \, \biggr) \, \textnormal{d}m + \int \frac{\lambda m^2}{\sqrt{(1 - m^2) (1 + \lambda m^2)}} \, \textnormal{d}m\Biggr] \\ \\
& = \frac{2}{\sqrt{q - v} \, (q + v)} \Biggl[\frac{\sqrt{(1 - m^2) (1 + \lambda m^2)}}{m} + E\bigl(m, \textnormal{i} \sqrt{\lambda} \, \bigr) - F\bigl(m, \textnormal{i} \sqrt{\lambda} \, \bigr)\Biggr] \, .
\end{split}
\end{equation}
Finally, substituting (\ref{Simpint}) and (\ref{32int}) into (\ref{s32int}), and subsequently --- after having employed the elliptic integral identities (\ref{ID1}) and (\ref{ID2}) --- (\ref{s32int}) into (\ref{I21}) yields
\begin{equation*}
\begin{split}
\frac{\mathcal{J}}{q \, \sqrt{2}} & = - \frac{\sqrt{v - q \sin{(u)}}}{3 \cos^3(u)} - \frac{q \tan{(u)}}{6 \sqrt{v - q \sin{(u)}}} + \frac{q}{6 \, \sqrt{q + v}} \Biggl[\frac{v}{\sqrt{v - q \sin{(u)}} \, \sqrt{q + v}} \, \sqrt{\frac{1 + \sin{(u)}}{1 - \sin{(u)}}} \\ \\
& \hspace{0.4cm} + \frac{v}{q - v} \, E\Biggl(\sqrt{\frac{v - q \sin{(u)}}{q \, [1 - \sin{(u)}]}}, \sqrt{\frac{2 q}{q + v}} \, \Biggr) + F\Biggl(\sqrt{\frac{v - q \sin{(u)}}{q \, [1 - \sin{(u)}]}}, \sqrt{\frac{2 q}{q + v}} \, \Biggr)\Biggr] \, .
\end{split}
\end{equation*}
In cylindrical coordinates, this function becomes
\begin{equation*} 
\begin{split}
& \frac{3 \mathcal{J}}{q} = - \frac{r^3 + q z}{\rho^2} + \frac{q}{\sqrt{\rho^2 + 2 q \, (1 + z/r)}} \Biggl[\frac{1}{\rho} \, \frac{\rho^2 + 2 q z/r}{\sqrt{\rho^2 + 2 q \, (1 + z/r)}} \, \sqrt{\frac{r + z}{r - z}} - \frac{\rho^2 + 2 q z/r}{\rho^2 - 2 q \, (1 - z/r)} \\ \\
& \times E\Biggl(\frac{\rho}{\sqrt{2 q \, (1 - z/r)}}, \, 2 \, \sqrt{\frac{q}{\rho^2 + 2 q \, (1 + z/r)}} \, \Biggr) + F\Biggl(\frac{\rho}{\sqrt{2 q \, (1 - z/r)}}, \, 2 \, \sqrt{\frac{q}{\rho^2 + 2 q \, (1 + z/r)}} \, \Biggr)\Biggr] \, .
\end{split}
\end{equation*}

\section*{Appendix B: Spherical Coordinate Representation of the Magnetic Field Solution} \label{AppB}

\noindent We express the general magnetic field solution (\ref{Bcomp}) for the inner heliosheath by means of a specific spherical coordinate system whose polar axis is aligned with the Sun's axis of rotation, which makes it more suitable for the description of the interplanetary magnetic field (see, e.g., \cite{Parker-1958}). This is necessary in order to formulate the corresponding initial-boundary conditions, as they establish the link between these two magnetic fields. We begin by deriving the relations between the cylindrical coordinates defined in (\ref{T0C}) and the spherical coordinates for the interplanetary region $(r', \vartheta', \phi') \in \mathbb{R}_{> 0} \times (0, \pi) \times [0, 2 \pi)$ with
\begin{equation} \label{ASC}
\begin{split}
& r' = \sqrt{X'^2 + Y'^2 + Z'^2} \, , \quad \vartheta' = \arccos{\biggl(\frac{Z'}{\sqrt{X'^2 + Y'^2 + Z'^2}}\biggr)} \, , \\ \\
& \textnormal{and} \quad \phi' = \textnormal{sgn}(Y') \biggl[\arccos{\biggl(\frac{X'}{\sqrt{X'^2 + Y'^2}}\biggr)} - \pi\biggr] + \pi \, , 
\end{split}
\end{equation}
where $(X', Y', Z') \in \mathbb{R}^3$ are Cartesian coordinates for which the orientation of the associated basis differs from the one employed in Section \ref{SecII} by a counterclockwise rotation about the $Y'$ axis through an angle of $\pi/2 \,\, \textnormal{rad}$. Accordingly, we have to perform the transformation 
\begin{equation*} 
\mathcal{R} \hspace{-0.05cm}:
\begin{cases}
\hspace{0.18cm} \mathbb{R} \times \mathbb{R} \times \mathbb{R} \rightarrow \mathbb{R} \times \mathbb{R} \times \mathbb{R} \vspace{0.2cm} \\ 
\, (X', Y', Z') \mapsto (X, Y, Z)
\end{cases} 
\end{equation*}
with
\begin{equation*} 
X = - Z' \, , \quad Y = Y' \, , \quad \textnormal{and} \quad Z = X' \, ,
\end{equation*}
and insert the latter coordinate relations together with the inverse of (\ref{ASC}) into (\ref{T0C}), resulting in
\begin{equation*} 
\begin{split}
& \rho = r' \sqrt{1 - \textnormal{si}\textnormal{n}^2{(\vartheta')} \cos^2(\phi')} \, , \quad z = r' \sin{(\vartheta')} \cos{(\phi')} \, , \\ \\
& \textnormal{and} \quad \varphi = \textnormal{sgn}(\phi' - \pi) \, \arccos{\biggl(\frac{\cos{(\vartheta')}}{\sqrt{1 - \textnormal{si}\textnormal{n}^2{(\vartheta')} \cos^2(\phi')}}\biggr)} + \pi \, .
\end{split}
\end{equation*}
The relations between the cylindrical unit base vectors and the spherical unit base vectors thus read 
\begin{equation*}
\begin{split}
\boldsymbol{e}_{\rho} & = \sqrt{1 - \textnormal{si}\textnormal{n}^2{(\vartheta')} \cos^2(\phi')} \,\, \boldsymbol{e}_{r'} + \frac{\sin{(\vartheta')} \cos{(\phi')}}{\sqrt{1 - \textnormal{si}\textnormal{n}^2{(\vartheta')} \cos^2(\phi')}} \, \bigl[- \cos{(\vartheta')} \cos{(\phi')} \, \boldsymbol{e}_{\vartheta'} + \sin{(\phi')} \, \boldsymbol{e}_{\phi'}\bigr] \\ \\
\boldsymbol{e}_z & = \sin{(\vartheta')} \cos{(\phi')} \, \boldsymbol{e}_{r'} + \cos{(\vartheta')} \cos{(\phi')} \, \boldsymbol{e}_{\vartheta'} - \sin{(\phi')} \, \boldsymbol{e}_{\phi'} \\ \\
\boldsymbol{e}_{\varphi} & = - \frac{1}{\sqrt{1 - \textnormal{si}\textnormal{n}^2{(\vartheta')} \cos^2(\phi')}} \, \bigl[\sin{(\phi')} \, \boldsymbol{e}_{\vartheta'} + \cos{(\vartheta')} \cos{(\phi')} \, \boldsymbol{e}_{\phi'}\bigr] \, .
\end{split}
\end{equation*}
Then, substituting these expressions as well as the solution (\ref{Bcomp}) into the invariance condition
\begin{equation*}
B_{\rho} \, \boldsymbol{e}_{\rho} + B_z \, \boldsymbol{e}_z + B_{\varphi} \, \boldsymbol{e}_{\varphi} = B_{r'} \, \boldsymbol{e}_{r'} + B_{\vartheta'} \, \boldsymbol{e}_{\vartheta'} + B_{\phi'} \, \boldsymbol{e}_{\phi'} \, ,
\end{equation*}
we can directly identify the desired spherical representation of the magnetic field components, yielding
\begin{subequations} \label{mfc} 
\begin{align}	
B_{r'} & = \biggl(\frac{1}{r'^2} - \frac{\sin{(\vartheta')} \cos{(\phi')}}{q}\biggr) [3 \mathcal{J} \mathcal{T}_1 + \mathcal{T}_2] - \frac{r' \sin{(\vartheta')} \cos{(\phi')} \, \mathcal{T}_1}{1 - \sin^2{(\vartheta')} \cos^2{(\phi')}}  \\ \nonumber \\ 
B_{\vartheta'} & = - \frac{\cos{(\vartheta')} \cos{(\phi')} \, [3 \mathcal{J} \mathcal{T}_1 + \mathcal{T}_2]}{q} - \frac{r' \cos{(\vartheta')} \cos{(\phi')} \, \mathcal{T}_1}{1 - \sin^2{(\vartheta')} \cos^2{(\phi')}} - r' \sin{(\phi')} \, \mathcal{T}_3 \\ \nonumber \\
B_{\phi'} & = \frac{\sin{(\phi')} \, [3 \mathcal{J} \mathcal{T}_1 + \mathcal{T}_2]}{q} + \frac{r' \sin{(\phi')} \, \mathcal{T}_1}{1 - \sin^2{(\vartheta')} \cos^2{(\phi')}} - r' \cos{(\vartheta')} \cos{(\phi')} \, \mathcal{T}_3 \, ,
\end{align}
\end{subequations}
where
\begin{equation*} 
\mathcal{T}_1 := \frac{\partial_{\chi} \psi}{u_0 q} + \partial_{\mu} \mathscr{C}_{\varphi} \, , \quad \mathcal{T}_2 := \partial_{\chi} \mathcal{H} - q \, \partial_w \mathscr{C}_{\varphi} \, , \quad \textnormal{and} \quad \mathcal{T}_3 := \frac{\partial_w \psi}{u_0} + \partial_{\mu} \mathcal{H} \, .
\end{equation*}

\end{appendix}

\vspace{-0.25cm}


\end{document}